\documentclass[twocolumn,showpacs,preprintnumbers,amsmath,amssymb]{revtex4}
\usepackage{graphics}
\usepackage[pdftex]{graphicx}

\begin{document}
\title{Pathways of bond topology transitions at the interface of silicon nanocrystals and amorphous silica matrix}
\author{D. E. Y{\i}lmaz}
\email{dundar@fen.bilkent.edu.tr}
\author{C. Bulutay}
\email{bulutay@fen.bilkent.edu.tr}
\affiliation{Department of Physics, Bilkent University, Ankara, 06800, Turkey}
\author{T. \c{C}a\u{g}{\i}n}
\email{cagin@chemail.tamu.edu}
\affiliation{Texas A$\&$M University, Artie McFerrin Department of Chemical Engineering,\\
Jack E. Brown Engineering Building, 3122 TAMU, College Station, TX 77843-3122, USA}
\begin{abstract}
The interface chemistry of silicon nanocrystals (NCs) embedded in amorphous oxide matrix 
is studied through molecular dynamics simulations with the chemical environment 
described by the reactive force field model. Our results indicate that the Si NC-oxide 
interface is more involved than the previously proposed schemes which were 
based on solely simple bridge or double bonds. We identify different types of 
three-coordinated oxygen complexes, previously not noted. The abundance and the 
charge distribution of each oxygen complex is determined as a function of the NC size as well as the 
transitions among them. 
The oxidation at the surface of NC induces tensile strain to Si-Si bonds which become significant only 
around the interface, while the inner core remains unstrained.
Unlike many earlier reports on the interface structure, we do not observe any double bonds. 
Furthermore, our simulations and analysis reveal that the interface bond topology evolves among different oxygen 
bridges through these three-coordinated oxygen complexes. 
\end{abstract}

\pacs{61.46.Hk, 68.35.Ct}

\maketitle
\section{Introduction}
After a long arduous effort, photoluminescence from silicon has been achieved from its 
nanocrystalline form~\cite{canham}. A critical debate, however, continues over the nature of the interface chemistry of silicon 
nanocrystals (Si-NCs) embedded in amorphous silica which has direct implications on the optical activity of the 
interface~\cite{wolkin,puzder,luppi,gatti,vasiliev}. Wolkin \textit{et al.} reported that the oxidation of porous 
silicon quantum dots results in a red shift in the photoluminescence (PL) spectra which indicates 
the importance of oxygen-related interface bond toplogy~\cite{wolkin}.
Along this line, Puzder and co-workers compared PL calculations of nanoclusters with different passivants and 
surface configurations and proposed  the main reason for the red shift to be double Si=O bonds \cite{puzder}.
 Countering this,  Luppi \textit{et al.} reported excitonic luminescence features caused by Si-O-Si bridge bonds 
 at the surface of silicon nanoclusters \cite{luppi}. As a supporting evidence for the latter, 
Gatti \textit{et al.} recently demonstrated  that Si-O-Si is the most stable isomer configuration 
\cite{gatti}. To reconcile, Vasiliev 
 \textit{et al.} claimed that bridge bonds and double bonds have similar effect on PL \cite{vasiliev}. 

All of the work cited above represent density functional theory (DFT)-based
 calculations with small Si clusters of less than 100 atoms surrounded by either passivants like 
hydrogen \cite{puzder,luppi,gatti} or oxygen \cite{wolkin,puzder}. 
But actual samples are profoundly different: the 
fabricated systems consist of Si-NC with a diameter larger than $1$~nm, embedded in 
\textit{amorphous} silica (a-SiO$_2$) matrix. Identifying this fact, Tu and Tersoff studied Si/a-SiO2 interface using a bond order dependent empirical potential and proposed that Si-O-Si bridges are 
the main blocks at the interface lowering the surface energy \cite{tersoff}. Using the same model potential 
Hadjisavvas \textit{et al.} studied Si-NCs embedded in a-SiO$_2$ and they also reported bridge 
bonds as the mechanism for lowering surface strain energy \cite{hadjisavvas_surface}. In our previous work on the formation and structure of Si NCs, we also 
observed that this model potential falls short to characterize the system accurately, especially the structure and chemistry at the interface \cite{dundar}. As a rigorous approach to interface structure and dynamics, Pasquarello \textit{et al.} used first-principles 
molecular dynamics (MD) technique to investigate planar Si/SiO$_2$ interface \cite{pasquarello_nature}. 
They observed that the oxygen atoms 
momentarily are bonded to three silicon atoms during oxidation process. The net effect of these threefold 
coordinated oxygen atoms during oxidation process was to expel silicon atoms out of interface. This is 
interpreted as a balancing process to decrease the increased density of Si/SiO$_2$ interface 
due to oxidation \cite{pasquarello_nature}. However, no substantial breakthrough was made over the last decade on this issue.
%
\section{Method}
%
In this work, we employ the reactive force field (ReaxFF) model developed by van Duin \textit{et al.} 
which improves Brenner's reactive bond order model \cite{brenner} to a level of accuracy and validity 
allowing molecular dynamics simulations of the full reaction pathways in bulk~\cite{adri}. The parameters for this force field were obtained from fitting to the results of ab initio calculations on relevant species as well as periodic boundary condition DFT-based calculations of various crystalline polymorphs of relevant materials. 
The ReaxFF calculates bond orders which is the measure of bond strength from local geometry. This allows realistic chemical environment such as over/under coordination and bond breaking/formation for large-scale (about 5000 atoms) MD simulations.

\begin{figure}
\includegraphics{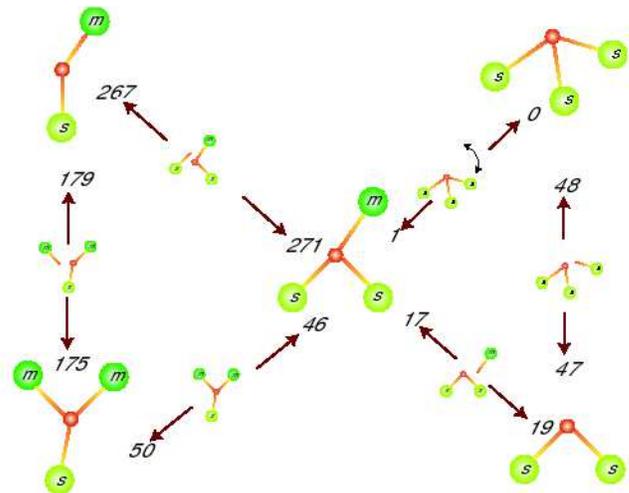}
\caption{(Color online) 
The transitions between different oxygen complexes bonded to the interface.
Dark green (dark gray) large spheres represent matrix silicon atoms, and the light green (light gray) large spheres represents surface silicon atoms of the NC, small red (dark gray) spheres represent oxygen atoms. The numbers indicate the number of transitions recorded in the simulation in each direction among the 
complexes for the NC of radius $13.4$~\AA. }
\label{fig:abbrevations}
\end{figure}

To facilitate our discussion regarding the surface-bonded oxygen complexes, we distinguish among 
three different types of silicon atoms. We label those silicon atoms with all silicon 
neighbors each with zero oxidation state as \textit{c}, to denote core silicon atom. 
Among the \textit{remaining} (non-\textit{c}) silicon atoms, those with at least one 
bond to \textit{c} are labeled as \textit{s}, denoting as a surface silicon atom. For further investigation of NC, we seperate core Si atoms into two sub-categories as inner-core and outer-core atoms: Among core Si atoms which have at least one surface Si neighbour categorized as outer-core Si atoms and rest of core Si atoms categorized as inner-core Si atoms. Finally, 
any other silicon atom is labeled as \textit{m}, denoting matrix silicon atom. 
Hence, a complex consisting of an oxygen atom bonded with two surface silicon atoms is 
labeled as \textit{ss}. The other oxygen complexes are \textit{sm}, \textit{ssm}, 
\textit{sss}, \textit{mms} as sketched in Fig.~\ref{fig:abbrevations} where the last 
three correspond to three-coordinated oxygen (3cO) atoms.

We use ReaxFF to represent the interactions in the model system. We start with a large simulation cell 
(box length $43$~\AA) of silica glass formed 
through a melting and quenching process used by one of us \cite{tahir1,tahir2} earlier to study silica glasses earlier. Next, similar 
to Hadjisavvas \textit{et al.} \cite{hadjisavvas_surface}, 
we delete all atoms within a predetermined radius to insert crystalline silicon to form NC.
In this way, we create NCs with radii ranging from $5.5$~\AA~to $16.7$~\AA~. For the largest NC we insert $967$ Si atoms into a spherical hole with radius $16.7$~\AA~created in amorphous matrix. Even for this case, the minimum distance between NC surface to simulation box face is about 
$5$~\AA~which can still accommodate the interface layer. We also pay special attention in the removal of spherical region so that the correct stoichiometry for the amorphous matrix is met. Thus, our amorphous matrix has two O atoms for every Si atom with a 
density of $2.17$~g/cm$^3$ which is the density of glass at room temperature and atmospheric pressure. 

We set periodic boundary conditions in all directions and while keeping NC at 100~K, we employ simulated annealing process to SiO$_2$ region to end up with an amorphous matrix free of artificial strain around the NC. Then we set 
whole systems' temperature to room temperature (300~K) and continue performing MD simulation for 75~ps to have thermal 
equilibrium between the two regions. We set MD simulation time step to 0.25~fs for all simulations. For every 62.5~fs time interval, we have record the configurations to analyze the transitions 
taking place between different bond topologies. The numbers attached to each arrow in 
Fig.~\ref{fig:abbrevations} indicate the total number of registered transitions 
during the simulation in that direction between the complexes for a representative NC 
of radius $13.4$~\AA. Almost balanced rates in opposite directions is an assurance of 
the attainment of the steady state in our simulation. Note that we do not observe any 
direct transition other than the paths indicated in Fig.~\ref{fig:abbrevations}. 
For instance, a direct transition of the complex 
\textit{ss} to \textit{sm} does not  take place, but it is possible through an intermediate 
transition over the \textit{ssm} which is a 3cO. We 
should also remark that the balanced transitions continue to take place after the 
steady state is attained which indicates that the interface bond topology is dynamics, i.e. not frozen.

\begin{figure}
\includegraphics{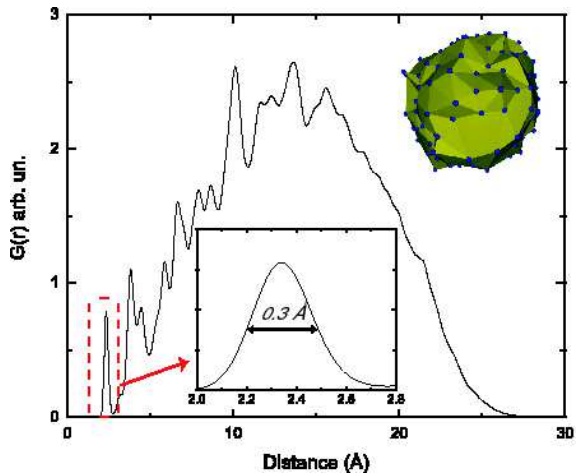}
\caption{(Color online) Radial distribution function of Si atoms in NC. Inset: First peak resembles
Si-Si bond length distribution centered around 2.34~\AA~with 0.3~\AA~FWHM value. Second peak's width resembles bond angle deviation. Inset: Representation of NC surface created with Delaunay Triangulation. Blue dots represents surface silicon atoms.}
\label{fig:rdfncsisi}
\end{figure}
To analyze Si-NC/a-SiO$_2$ interface, we construct NC surface using Delaunay
triangulation scheme (Fig.~\ref{fig:rdfncsisi} inset) \cite{delaunay}. 
In two dimensions a Delaunay triangulation of a set of points corresponds to
defining triangles such that no point in the set is inside the
circumcircle of any triangle. In three dimensions triangles extend to
tetrahedra and circumcircles become circumspheres. Since our NCs are
nearly spherical in shape, we triangulate projection points of surface
Si atoms onto the unit sphere. Hence, we apply Delaunay triangulation over the
two dimensional $\theta$-$\phi$ plane. We can then create surfaces using this triangulation. 
This surface enables us to calculate every atoms' distance to NC surface.
By this means we plot various data such as charge, bond order etc. with respect to distance 
to the surface, to extract information about surface chemistry of Si-NC embedded in amorphous matrix. 
%
\section{Results and Discussions}
\begin{figure}
\includegraphics{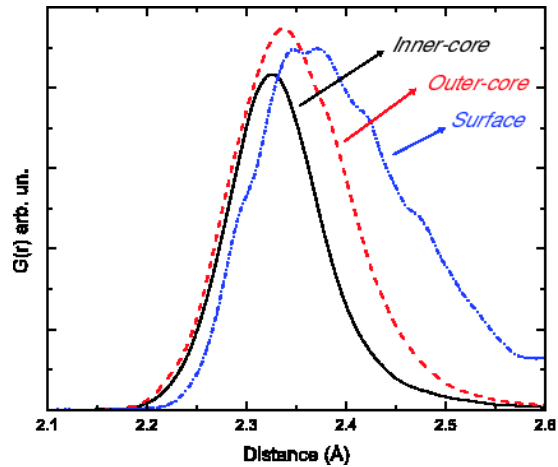}
\caption{ (Color online) Bond distance probability distribution of Si atoms in NC. Solid line represents
inner core Si-Si bonds, dashed line outer core Si-Si bonds, and dotted line represents
the surface Si-Si bonds.}
\label{fig:detailedrdf}
\end{figure}

\begin{figure}
\includegraphics{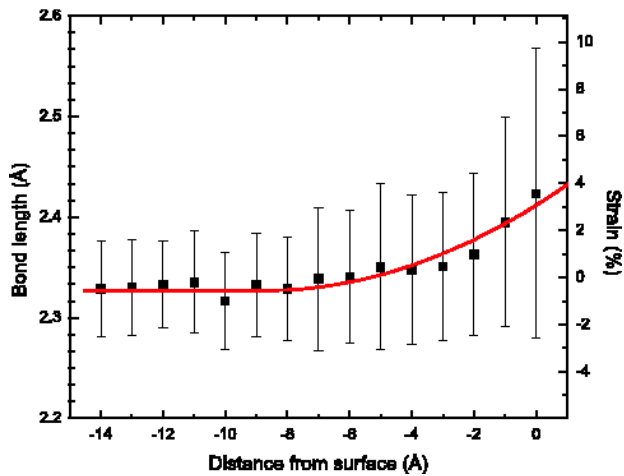}
\caption{(Color online)  Variation of Si-Si bond length averages (calculated over 1 A wide bins) as a function of distance from the NC surface -which is defined by Delaunay tesselation.  The solid line is a fit to the data to guide the eye.}
\label{fig:strain}
\end{figure}

%
For different NC radii we observe similar trends in bond order distribution, average charges etc., therefore, we present 
only the figures of the system for a typical NC of radius $13.4$~\AA. 
A useful data to elucidate the structure of these systems is the radial distribution function (RDF).  
In Fig.~\ref{fig:rdfncsisi} we present RDF of NC atoms only, where the first broad peak centered around 2.34~\AA~with a 0.3~\AA~full width at half maximum value (FWHM), represents Si-Si bond length distribution in NC (Fig.~\ref{fig:rdfncsisi} inset). The maximum extent of the NC can also be 
read from same plot at about 27~\AA, where the RDF goes to zero. Observation of a broad first peak at Fig~\ref{fig:rdfncsisi} demands further investigation of Si-Si RDF of NC atoms. For this purpose, in Fig.~\ref{fig:detailedrdf} we present bond length probability distributions (akin to RDF) for three categories of Si atoms: inner core (with no bonds to surface atoms), outer core (bonded to surface) and surface NC atoms. We observe from Fig.~\ref{fig:detailedrdf} that Si-Si bond lengths in the inner core are centered around the equilibrium value and have a narrow width due mainly to thermal vibrations, whereas the bond length distributions of outer core and surface atoms have increasing shift for the most probable bond length values and broader widths. These shifts and particularly the increase in distribution widths cannot be attributed to thermal broadening. Taken together these two observations is a clear indication of increasing strain as a function of distance from the center of the NC.  To further investigate this deviation of Si-Si bonds from crystalline Si, in Fig.~\ref{fig:strain} we present bond length distribution with respect to distance to NC surface averaged over 2~ps of simulation time after the steady state is reached. This figure illustrates the gradual development of radial strain from the center to NC surface.
These observations show clearly that oxidation at the surface of NC results in a tensile strain at Si-Si bonds which becomes significant only around the interface, while keeping the inner core almost unstrained. This tensile strain in the NC agrees with previous measurement of Hofmeister \textit{et al.} \cite{Hofmeister}. 

\begin{figure}
\includegraphics{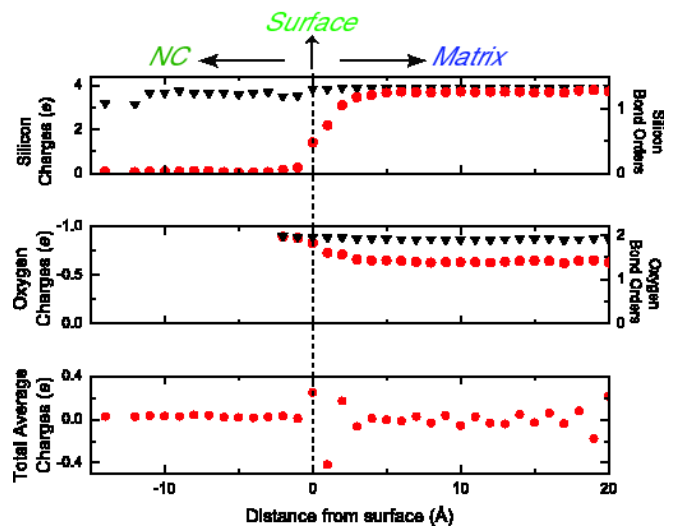}
\caption{ (Color online) Top: Silicon bond orders (triangles) and charges  (circles) as a function of distance from surface of NC;
Middle: Oxygen bond orders and charges as a function of distance from the surface of NC;
Bottom: Total average charge as a function of distance  from the surface of NC.  The averaging
bin width is 1~\AA.}
\label{fig:panel3}
\end{figure}

Another consequence of this tensile strain is that the total bond orders of core-NC atoms are somewhat smaller than those of oxide-Si's as seen in Fig.~\ref{fig:panel3}. In the same figure we also show the calculated net charges using electron equilibration method \cite{eem}. Nearly zero net charges of the core-Si atoms reflects the covalent type of bonding well within the NC. The bonding becomes increasingly ionic away from the NC core as observed by the charges of Si atoms which reach the value of 1.3$e$ at the oxide region (Fig.~\ref{fig:panel3}). As a result, the positive charges of surface-Si atoms form a shell at the surface of NC. This observation is similar with those 
obtained with DFT calculations \cite{Krollstatsol,Kroll}. 
 On the other hand, negative charges of oxygen atoms bonded to surface form another shell that 
 enclose NC and finally total average charges approach to zero within the oxide region.
 In Fig.~\ref{fig:panel3} we also observe that the magnitude of average charges of oxygen atoms 
which are bonded to surface are greater than those in the matrix. But, the bond orders are nearly 
the same. This is due to existence of 3cO atoms bonded to surface. 
Note that the average bond order of oxygen atoms which are bonded to surface is about two 
(cf., Fig.~\ref{fig:panel3}). 
Thus, those oxygen atoms form three partial bonds, 
two strong and one weak bond. 
Finally, we would like to note that unlike many others~\cite{wolkin,puzder,vasiliev}, we do not 
observe any double bonds. 

\begin{figure}
\includegraphics{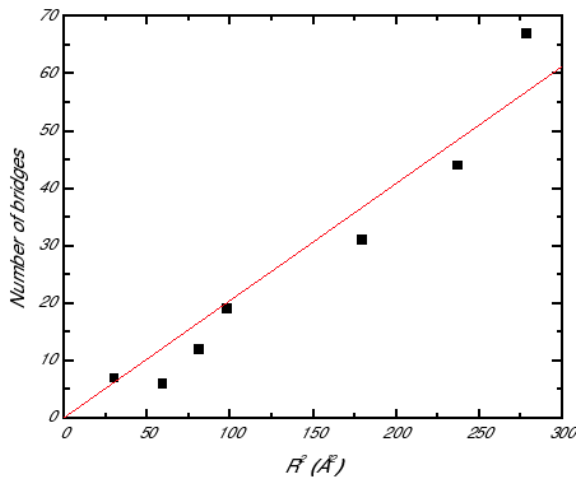}
\caption{(Color online) The number of bridges at the Si-NC surface vs radius squared. 
The line is a linear fit to data.}
\label{fig:nobridgesr2}
\end{figure}

\begin{table*}
\caption{Statistical results of atom charges and numbers, $N$, for all NC diameters, 
$D_{\mbox{NC}}$  considered. 
Abbreviations for atom types are explained in Fig.~\ref{fig:abbrevations}. Charges are in the units of electronic charge 
and the angles $\theta$, are in degrees.}
\begin{ruledtabular} 
 \begin{tabular}
{cccc ccc cc ccc cc cc c}
 $D_{\mbox{\begin{scriptsize}NC\end{scriptsize}}}$ &$N_c$ &$N_s$&$N_O$
&\multicolumn{3}{c}{\textit{ss}}&\multicolumn{2}{c}{\textit{sm}}&\multicolumn{3}{c}{\textit{ssm}}
&\multicolumn{2}{c}{\textit{sss}}&\multicolumn{2}{c}{\textit{mms}}&\\
(\AA)&&&&$N_{ss}$ &charge&$\theta_{ss}$&$N_{ss}$ &charge
&$N_{ssm}$ &charge&$\theta_{ss}$&$N_{sss}$ &charge&$N_{mms}$ &charge&\\
\cline{5-7}\cline{8-9} \cline{10-12}\cline{13-14}\cline{15-16}
 11.0 &  10  &  25  &  32  &  2 & -0.88  & 169.7 &  24 & -0.76 &  5  & -0.74  &  97.0 &  0 &   ---&  1 & -0.81& \\
 15.4 &  42  &  42  &  59  &  5 & -0.77  & 136.1 &  51 & -0.74 &  1  & -0.73  &  82.7 &  0 &   ---&  2 & -0.77& \\
 18.0 &  83  &  62  &  77  & 11 & -0.82  & 120.5 &  56 & -0.74 &  1  & -0.83  & 141.2 &  0 &   ---&  9 & -0.79& \\  
 19.8 & 114  &  76  &  82  & 14 & -0.82  & 139.3 &  49 & -0.73 &  4  & -0.81  & 123.1 &  1 & -0.85& 14 & -0.79& \\
 26.8 & 353  & 143  & 170  & 20 & -0.81  & 118.0 & 123 & -0.74 & 11  & -0.79  & 120.9 &  0 &   ---& 15 & -0.76& \\
 30.8 & 558  & 203  & 243  & 35 & -0.83  & 123.9 & 159 & -0.74 & 10  & -0.80  & 115.9 &  2 & -0.80& 34 & -0.79& \\
 33.4 & 718  & 238  & 268  & 44 & -0.83  & 123.0 & 179 & -0.75 & 18  & -0.80  & 125.7 &  4 & -0.84& 23 & -0.78& \\
 \end{tabular}
\end{ruledtabular}
\label{table:statatom}
\end{table*}
The occurrence of 3cO has been noted by a number of groups. Pasquarello showed that the bistable 
$E'_1$ defect of $\alpha$-quartz structure may lead to 3cO as a metastable state as well 
as Si-Si dimer bond  and calculated the energy of the former to be higher than the latter. 
Pasquarello proposed that 3cO acts as an \textit{intermediate metastable state} during structural 
relaxations at the interface \cite{pasquarello_surf}. Similarly Boero \textit{et al.} 
observed 3cO atoms in their \textit{ab initio} calculations \cite{boero} and reported 
this feature as a metastable state.
 In Table~\ref{table:statatom} we present the collected statistical data at 
 the end of the simulation of 75~ps. For all oxygen complexes, the number of 
 bridges, average charges of bridge oxygens, and the average bridge 
 angles for \textit{s}-O-\textit{s} are tabulated. 
 We observe in Table~\ref{table:statatom} that the number of 
 \textit{sss} complexes is very small due to narrow bond angle requirement of 
 this configuration. For the remaining 3cO complexes, \textit{ssm} and \textit{mms}, 
 their percentages are seen to increase with curvature.
 This can explain the fact that other studies \cite{pasquarello_surf,boero} which 
 have identified the 3cO complexes as metastable were based on the \textit{planar} 
 Si/SiO$_2$ interfaces. So, this is an indication of the importance of 
 curvature in the stability of 3cO complexes.
 Hence, as one would expect, there is a linear relation between the 
 total number of bridges with surface area as indicated in Fig.~\ref{fig:nobridgesr2}. This 
 finding is supported by Kroll \textit{et al.} who reported  3 and 
 33 such bridges for Si-NC with radii $4.0$~\AA{} and $7.0$~\AA, respectively~\cite{Krollstatsol}.
%
\section{Conclusions}
%
In conclusion, the realistic chemical environment provided by reactive force field model
enables us to understand the bond topology of Si-NC/a-SiO$_2$ interface and its internal dynamics. 
Particularly, it reveals that there are different types of oxygen complexes at the Si-NC 
surface some of which contain 3cO complexes whereas there are no double bonds. 
The curvature has a positive effect on the occurence of 3cO.
The relative abundance of different complexes and their charge and geometrical 
characteristics are extracted. 
The inner core is observed to be almost unstrained while the outer core and the interface region
of the NC are under increasing strain up to about a few per cents.
In general, our work clearly shows that the Si NC-oxide interface 
is more complicated than the previously proposed schemes which were based on solely  
simple bridge and double bonds. 
The provided information here paves the way to construct realistic Monte Carlo moves for the
simulation of large-scale silicon nanostructures embedded in oxide matrix.
\begin{acknowledgments}
This work has been supported by the European FP6 Project SEMINANO with the Contract
No. NMP4 CT2004 505285, by the Turkish Scientific and Technical Council T\"UB\.ITAK with the Project
No. 106T048. The visit of Tahir  \c{C}a\u{g}{\i}n to Bilkent University was facilitated by the 
T\"UB\.ITAK B\.IDEB-2221 programme. The computational resources are supplied in 
part by T\"{U}B\.ITAK through TR-Grid e-Infrastructure Project.
\end{acknowledgments}

\end{document}